\documentclass[authoryear,5p,times,twocolumn]{elsarticle}
\usepackage{graphicx}          
\usepackage{natbib}        
\usepackage[tight,footnotesize]{subfigure}
\usepackage{amssymb}
\usepackage{amsmath,calc}
\usepackage{color}
\usepackage{epstopdf}
\usepackage{float}

\usepackage{animate}
\usepackage{enumerate}

\newcommand{\Lo}{\mathcal{L}}

\newcommand{\Co}{\mathcal{C}}
\newcommand{\Po}{\mathcal{P}}
\newcommand{\Ao}{\mathcal{A}}

\newcommand{\Real}{\mathbb{R}}

\newcommand{\dt}{\partial t}
\newcommand{\fb}{f_\partial}
\newcommand{\eb}{e_\partial}
\newcommand{\ze}{\zeta}

\newtheorem{theo}{Theorem}
\newtheorem{remark}[theo]{Remark}
\newtheorem{definition}[theo]{Definition}
\newtheorem{assumption}[theo]{Assumption}

\newtheorem{proposition}[theo]{Proposition}
\newtheorem{corollary}[theo]{Corollary}
\newtheorem{lemma}[theo]{Lemma}
\newtheorem{example}[theo]{Example}
\newenvironment{Proof}[1][Proof]{\textbf{#1.} }{\ \rule{0.5em}{0.5em}}

\journal{Automatica}

\begin{document}

\begin{frontmatter}

\title{Observer-based boundary control of distributed port-Hamiltonian systems} 
\author[First]{Jesús Toledo} 
\author[First]{Yongxin Wu}
\author[Second]{Héctor Ramírez}
\author[First]{Yann Le Gorrec}

\address[First]{FEMTO-ST Institute, 
Univ. Bourgogne Franche-Comt\'e, ENSMM, CNRS,
24 rue Savary, F-25000 Besan\c{c}on, France. jesus.toledo@femto-st.fr, yongxin.wu@femto-st.fr, legorrec@femto-st.fr. }
\address[Second]{Department of Electronic Engineering, Universidad Técnica Federico Santa María, Avenida España 1680, Valparaiso, Chile. hector.ramireze@usm.cl}
    
\begin{abstract}                          
An {observer-based} boundary controller for infinite-dimensional port-Hamiltonian systems defined on 1D spatial domains is proposed. The design is based on an early-lumping approach in which a finite-dimensional approximation of the infinite-dimensional system {derived by spatial discretization} is used to design the observer and the controller. {As long as the finite-dimensional approximation approaches the infinite-dimensional model, the performances also do.} The main contribution is a constructive method which guarantees that the interconnection between the controller and the infinite-dimensional system is asymptotically stable. A Timoshenko beam model has been used to illustrate the approach.
\end{abstract}
\end{frontmatter}

\section{Introduction}
{Boundary control systems (BCS) \citep{Fattorini1968JournalBoundary} are} a class of control systems where the dynamics {are} described by partial differential equations (PDEs) with actuation and measurement situated at the boundaries of the spatial domain. Motivated by technological advances, these type of systems have been of great interest for engineers and mathematicians during the last decades since a large class of physical processes can be represented as BCS. This is for instance the case of beams and waves in mechanical systems, heat bars and bed reactors in chemical systems or telegraph equations in electronic systems, among others \citep{Curtain2012BookAnIntroduction}.

Recently, the control of BCS has been addressed by using the framework of port-Hamiltonian systems (PHS)  \citep{Van2002JournalHamiltonian,LeGorrec2005JournalDirac}. Boundary controlled PHS (BC-PHS) are an extension of the Hamiltonian formulation of mechanical systems to open multi-physical systems \citep{Duindam2009BookModeling}. This formalism has been proven to be particularly suitable for the modeling and control of complex physical systems, such as systems described by infinite-dimensional or non-linear models. The stability, stabilization and control synthesis of BC-PHS have been addressed in \citep{Villegas2009JournalExponential,Ramirez2014JournalExponential,Augner2014JournalStability,Macchelli2017JournalOnTheSynthesis,Ramirez2017JournalStabilization}. More recently, the framework has been extended to deal with robust and adaptive regulation \citep{Macchelli2018JournalDissipativity,Humaloja2018JournalRobust}.

In the case of observer-based control design there are generally two approaches. The first one is the late-lumping approach in which the observer is designed from the infinite-dimensional systems \citep{Guo2007JournalTheStabilization,Meurer2013JournalOnTheExtended}. The main problem comes from the infinite-dimensional aspect of the controller structure that needs to be reduced for practical and real-time implementation. The second one is the early-lumping approach. In this case, the system is first approximated and a finite-dimensional observer is implemented on the reduced order system. The main drawback is the spillover effect induced by the use of a reduced order controller on the infinite-dimensional system, leading to high-frequency mode destabilization \citep{Bontsema1988JournalAnote}.

The main result of this paper is the proposition of a systematic synthesis method for observer-based boundary controller design for BC-PHS defined on one dimensional spatial domains. A finite-dimensional PHS approximation of the BC-PHS is used to design a strictly positive real PHS {{observer-based} state feedback}. {This approximation is considered precise enough such that in the frequency range of interest the modes of the approximated model are close enough to the one of the original system.}

The observer is then used to compute the boundary control law for the infinite-dimensional system. Using the passivity properties of power preserving interconnection of PHS it is then possible to guarantee the asymptotic stability of the closed-loop system.  The controller hence allows to assign the low-frequency modes while guaranteeing stable high-frequency modes, avoiding spillover effects.  
This paper is organized as follows: Section \ref{Section:BackGround} gives a background on port-Hamiltonian systems. Section \ref{Section:MainResults} gives the main result of this work. Section \ref{Section:Example} presents a numerical example, namely a boundary actuated one-dimensional Timoshenko beam.

\section{Background on port-Hamiltonian systems}\label{Section:BackGround}
\subsection{Some notation}
In this paper, $M_n(\mathbb{R})$ denotes {the space of real $n \times n$ matrices } and $I$ denotes the identity matrix of appropriate dimensions. By $\langle \cdot , \cdot \rangle_{L_2}$ or only $\langle \cdot , \cdot \rangle$ we denote the standard inner product on $L_2(a,b;\mathbb{R}^n)$ and the Sobolev space of order $p$ is denoted by $H^p(a,b;\mathbb{R}^n)$. A detailed description of the class of boundary control systems under consideration can be found in \citep{LeGorrec2005JournalDirac,Jacob2012BookLinear}. In the next section, we recall some basic properties of this class of systems.

\subsection{Boundary controlled port-Hamiltonian systems}
The class of one-dimensional PDEs under study, with inputs and outputs at the spatial boundaries, is given by the following set of equations
\begin{equation}\label{Sys:Plant}
\Po\begin{cases}
\dfrac{\partial z}{\dt}(\ze,t) = P_1 \dfrac{\partial}{\partial \ze} (\Lo(\zeta) z(\ze,t)) +(P_0-G_0)\Lo(\ze) z(\ze,t), \\
\,\,\,\, u(t)  = W \bigl(\begin{smallmatrix}
\fb (t) \\
\eb (t)
\end{smallmatrix}\bigr), \,\,\,\,\,\,\,\, z(\ze,0) = z_0(\ze),  \\
\,\,\,\, y(t)  = \widetilde{W} \bigl(\begin{smallmatrix}
\fb (t) \\
\eb (t)
\end{smallmatrix}\bigr), \,\,\,\,\,\,\,\, t \geq 0, \,\, \ze \in [a,b],
\end{cases}
\end{equation}
with $t>0$ the time variable and {$\ze \in [a,b]$} the 1D spatial coordinate, and  $z(\ze,t) \in \mathbb{R}^n$ the state variable. $P_1 = P_1 ^T  \in M_n(\mathbb{R})$ is a non-singular matrix, $P_0 = -P_0 ^T  \in M_n(\mathbb{R})$, $G_0 = G_0 ^T \geq 0  \in M_n(\mathbb{R})$, $\Lo (\cdot) \in M_n(L_2(a,b))$ is a bounded and continuously differentiable matrix-valued function satisfying for all $\ze \in [a,b]$, $\Lo (\ze)=\Lo ^T (\ze)$ and $mI < \Lo (\ze)< MI$ with $M>m>0$ both scalars independent on $\ze$. The state space is $Z = L_2(a,b;\mathbb{R}^n)$ with inner product $\langle z_1, z_2 \rangle_\Lo = \langle z_1, \Lo z_2 \rangle$ and norm $\Vert z \Vert^2 _\Lo = \langle z, z \rangle_\Lo$, hence $Z$ is a Hilbert space. The norm $\Vert\cdot \Vert ^2 _\Lo $ is usually proportional to the stored energy of the system, hence $z(\ze,t)$ is called energy variable and $\Lo(\ze)z(\ze,t)$ is called co-energy variable.  For simplicity, we write $z$ and $\Lo z$  instead of $z(\ze,t)$ and $\Lo (\ze)z(\ze,t)$ unless otherwise stated and arguments of dependent variables may be omitted.

\begin{definition}
Let $\Lo z \in H^1(a,b;\mathbb{R}^n)$. Then, the boundary port variables associated with (\ref{Sys:Plant}) are the vectors $\fb$ and $\eb \in \Real ^n$, defined by
\begin{equation}
\begin{pmatrix}
\fb (t) \\
\eb (t)
\end{pmatrix}=\frac{1}{\sqrt{2}}\begin{pmatrix}
P_1 & -P_1 \\
I & I
\end{pmatrix}\begin{pmatrix}
\Lo (b) z(b,t) \\
\Lo (a) z(a,t) \\
\end{pmatrix}.
\end{equation}
\end{definition}

Note that, the port-variables are nothing else than a linear combination of the boundary variables. We also define the matrix $\Sigma \in M_{2n}(\Real)$ as follows
\begin{equation}\label{Eq:Sigma}
\Sigma = \begin{pmatrix}
0 & I \\
I & 0
\end{pmatrix}.
\end{equation}

The following theorem ensures the existence and uniqueness of solutions of (\ref{Sys:Plant}).
\begin{theo}\label{Theorem:BCInputs}
\citep{LeGorrec2005JournalDirac} Let $W$ be a $n \times 2n$ real matrix. If $W$ has full rank and satisfies $W \Sigma W^T \geq 0$ then the system (\ref{Sys:Plant}) with input
\begin{equation}
u(t) = W \begin{pmatrix}
\fb (t) \\
\eb (t)
\end{pmatrix}
\end{equation} 
is a BCS on $Z$. Furthermore, the operator $\Ao z = P_1 \frac{\partial}{\partial \zeta}(\Lo z)+ (P_0-G_0) \Lo z$ with domain
\begin{equation*}
D(\Ao) = \left \lbrace \Lo z \in H^1(a,b;\Real ^n) \Bigg \vert \begin{pmatrix}
\fb (t) \\
\eb (t)
\end{pmatrix} \in ker \, W \right \rbrace
\end{equation*}
generates a contraction semigroup on $Z$.
\end{theo}
Let $\widetilde{W}$ be a full rank matrix of size $n \times 2n$ with $[\begin{smallmatrix}
W ^T &
\widetilde{W} ^T
\end{smallmatrix}]^T$ invertible and $P_{W,\widetilde{W}}$ given by
\begin{equation}
P_{W,\widetilde{W}} = \begin{pmatrix}
W \Sigma W^T & W \Sigma \widetilde{W}^T \\
\widetilde{W} \Sigma W^T & \widetilde{W} \Sigma \widetilde{W}^T
\end{pmatrix}^{-1}.
\end{equation}
Define the output of the system as the linear mapping {$\Co: \Lo ^{-1} H^1(a,b;\Real^n)\longrightarrow \Real^n$ }
\begin{equation}
y(t) = \Co z(\ze,t) = \widetilde{W}\begin{pmatrix}
\fb (t) \\ \eb (t)
\end{pmatrix}
\end{equation}
Then, for {$u\in C^2(0,\infty;\Real^n)$}, $\Lo z(\ze,0)\in H^1(a,b;\Real ^n)$ and $u(0)= W \begin{pmatrix}
\fb (0)\\ \eb (0)
\end{pmatrix}$ the following balance equation is satisfied
\begin{equation}\label{Eq:Energy}
\frac{1}{2}\frac{d}{dt} \Vert z(\ze,t) \Vert _\Lo ^2 = \frac{1}{2} \begin{pmatrix}
u(t)\\
y(t)
\end{pmatrix}^T P_{W,\widetilde{W}}\begin{pmatrix}
u(t)\\
y(t)
\end{pmatrix}
\end{equation}
\begin{remark}
The matrix 
\begin{equation}
\begin{pmatrix}
W \\
\widetilde{W}
\end{pmatrix}\Sigma \begin{pmatrix}
W \\
\widetilde{W}
\end{pmatrix}^T = \begin{pmatrix}
W \Sigma W^T & W \Sigma \widetilde{W}^T \\
\widetilde{W} \Sigma W^T & \widetilde{W} \Sigma \widetilde{W}^T
\end{pmatrix}
\end{equation}
is invertible if and only if {$[\begin{smallmatrix}
W^T &
\widetilde{W}^T
\end{smallmatrix}]^T$} is invertible. 
\end{remark}

In this work, we shall consider an early-lumping approach, i.e., the controller and observer are designed on a finite-dimensional approximation of (\ref{Sys:Plant}). The following assumption is considered 
\begin{assumption}\label{Assumption:pHDiscretization}
There exists the following finite-dimensional approximation of (\ref{Sys:Plant})
\begin{equation}\label{Sys:DiscretizedPlant}
 P \begin{cases}
\dot{x}(t) &= (J-R)Q x(t) + B u(t) \\
y(t) & = B^T Q x(t) \\
\end{cases}
\end{equation}
where $x \in \Real ^{n_c}$ with $n_c$ given by the order of the approximation, $J=-J^T$, $R=R^T \geq 0 $, $Q=Q^T >0$ all of them in $M_{n_c}(\mathbb{R})$ and $B \in \Real^{n_c \times n}$. Furthermore we assume (\ref{Sys:DiscretizedPlant}) to be controllable and observable. For simplicity, we shall define $A = (J-R)Q$ and $C = B^T Q$ and we will refer to the system $(A,B,C)$ as the approximated model of (\ref{Sys:Plant}).
\end{assumption}

\begin{remark}
Approximation schemes which preserve the port-Hamiltonian structure of the original system using mixed finite elements or finite differences on staggered grids for instance, can be found in  \citep{Seslija2012JournalDiscrete,Trenchant2018JournalFinite}. {The achievable closed-loop performances depend on the quality of the approximated model. The order of the approximation $n_c$ has then to be chosen large enough such that in the frequency range of interest the approximated system poles behave similar to the original ones.}
\end{remark}

\section{The observer-based controller}\label{Section:MainResults}
The main objective of this work is to design {a finite-dimensional controller} that achieves some desired performances on the finite-dimensional system \eqref{Sys:DiscretizedPlant} while ensuring closed-loop stability when applied to the infinite-dimensional system (\ref{Sys:Plant}). The considered controller is an {observer-based} state feedback
\begin{equation}\label{Eq:ControlLaw}
u(t) = r(t)-K\hat{x}(t)
\end{equation}
where $\hat{x}\in \Real ^{n_c}$, $r\in \Real ^{n}$ {and} the Luenberger observer
\begin{equation}\label{Eq:LuenbergerObserver}
\dot{\hat{x}}(t)=A\hat{x}(t)+Bu(t)+L(y(t)-C\hat{x}(t))
\end{equation}
with matrices $K\in \Real ^{n \times n_c}$, $L\in \Real ^{n_c \times n}$ to be designed and $(A,B,C)$ defined in (\ref{Sys:DiscretizedPlant}). Note that, $n_c$ is the size of the observer given by the chosen discretization scheme and $n$ is the number of boundary variables.

Several issues can arise when using an early-lumping approach to design the control, the most critical one being the loss of stability when the controller is applied on the infinite-dimensional system. It is known as the spillover effect \citep{Bontsema1988JournalAnote}. Consider the following illustrative example.

\begin{example} Consider the {1D} wave equation with unitary parameters and Neumann boundary control. The system can be written (see \cite{Jacob2012BookLinear} for more details) in the form (\ref{Sys:Plant}) with $$P_1=\left[ \begin{array}{cc}0 & 1 \\ 1 & 0 \end{array}\right], P_0=G_0=0, \Lo=I_2.$$
This model is discretized by using finite differences on staggered grids in order to preserve the structure of the system. Consider $n_c =59$ elements for the discretization. $G_0=0$ implies $R=0$ thus all the eigenvalues of $A$ are on the imaginary axis as shown in Figure \ref{Fig:ClosedLoopEigenvalues2} (a). $(A,B)$ is controllable and $(A,C)$ observable, hence $K$ and $L$ can be designed such that $A_K=A-BK$ and $A_L=A-LC$ are Hurwitz. Using for instance the {Linear Quadratic Regulator (LQR)} method the closed-loop eigenvalues can be assigned as in {Figure \ref{Fig:ClosedLoopEigenvalues2} (a)}.
                                 
The question that naturally arises is if the same control law, i.e. the same choice of matrices $K$ and $L$, preserves the stability when applied on the infinite-dimensional system. The answer in general is no. In this particular case for instance, when increasing the order of the discretized model to $n_c=67$, the closed-loop system turns unstable as shown in Figure \ref{Fig:ClosedLoopEigenvalues2} (b).
\begin{figure}[!htbp]
\begin{center}
\includegraphics[width=0.48\textwidth]{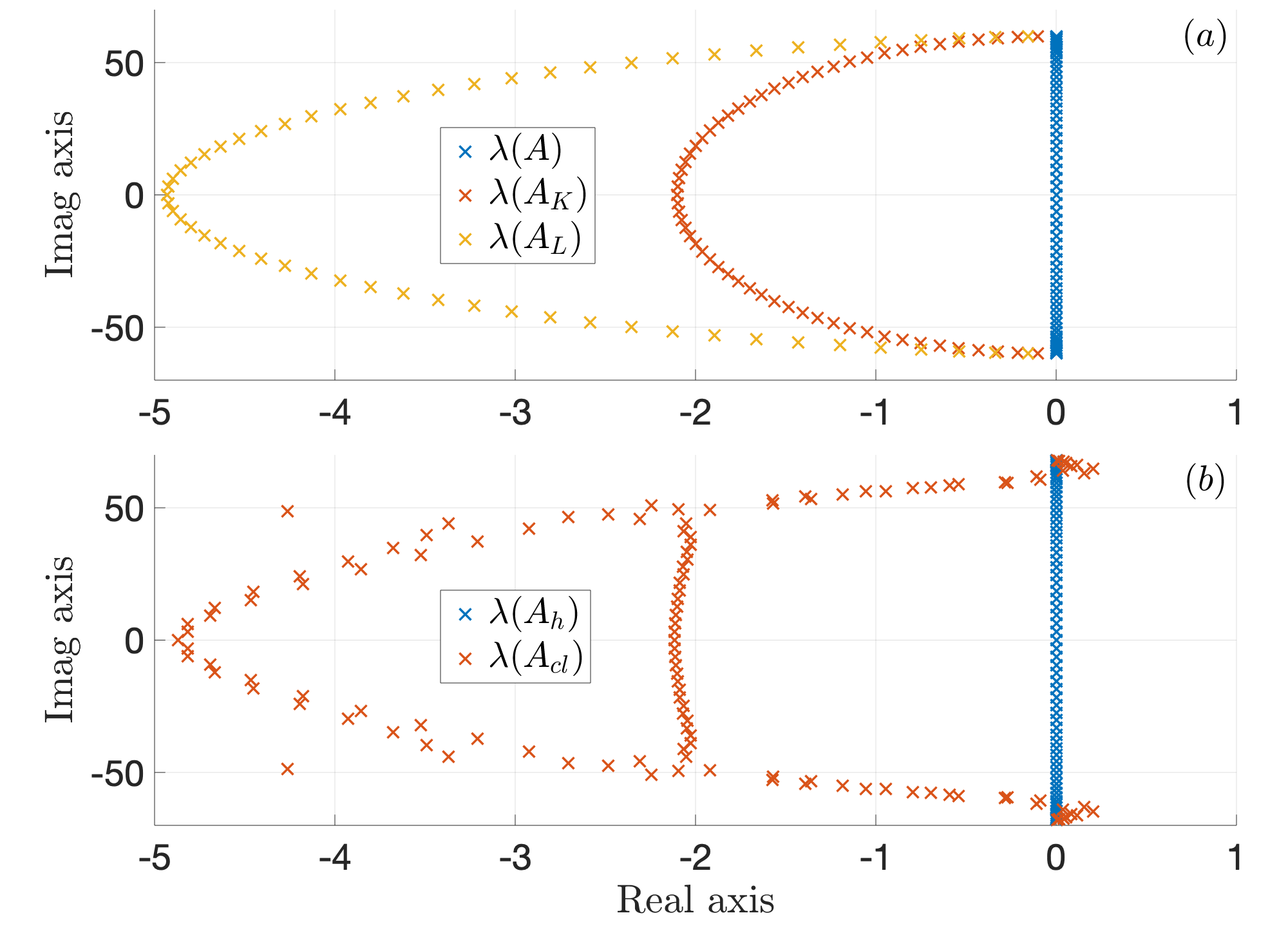}    
\caption{(a): $\lambda (A)$: discretized model eigenvalues with $n_c=59$, $\lambda (A_K)$: $A-BK$ eigenvalues and $\lambda (A_L)$: $A-LC$ eigenvalues. (b): $\lambda (A_h)$: discretized model eigenvalues with $n_c=67$, $\lambda (A_{cl})$: closed-loop eigenvalues.}  
\label{Fig:ClosedLoopEigenvalues2}                                 
\end{center}
\end{figure}   
\end{example} 
In what follows we start from the achievable closed-loop performances on the finite-dimensional system {\it i.e.} an appropriate choice of $K$, and design the observer gain such that the Luenberger observer (\ref{Eq:LuenbergerObserver}) is a strictly positive real PHS. Then we show that since (\ref{Eq:ControlLaw}) corresponds to a power preserving interconnection between the infinite-dimensional system and the dynamic boundary controller, the closed-loop system is asymptotically stable. 

\subsection{Some technical results}

Before presenting the main result we give the following definitions, lemma, corollary and theorem which are instrumental in the proof. 

\begin{definition}
A $n \times n$ transfer matrix $G(s)$ is positive real (PR) if $G(s)+G^T (\bar{s}) \geq 0$ for all $s$ such that $Re(s)>0$.
\end{definition}

\begin{definition}
A $n \times n$ transfer matrix $G(s)$ is strictly positive real (SPR) if there exists a scalar $\varepsilon >0$ such that $G(s-\varepsilon)$ is PR.
\end{definition}

\begin{lemma} (Lefschetz-Kalman-Yakubovich) \label{Lemma:LKY} \citep{Tao1988JournalStrictly}  Assume for the system $(A,B,C,D)$ that $(A,B)$ is controllable and $(A,C)$ is observable. Then, the transfer matrix $G(s)=C(sI-A)^{-1}B+D$ is SPR if and only if there exist real matrices $P=P^T >0$, $\Gamma$, {$W_1$ and a scalar $\varepsilon >0$ such that
\begin{subequations}
\begin{align}
PA + A^T P &= - \Gamma^T \Gamma - \varepsilon P \label{Eq:Lemma1} \\
C-B^T P &= W_1^T \Gamma \label{Eq:Lemma2} \\
D+D^T &= W_1^T W_1  \label{Eq:Lemma3}
\end{align}
\end{subequations}}
\end{lemma}

\begin{corollary}\label{Corollary:PH_SPR}
The system $(A,B,C,D)$ with $A=(J-R)Q$, $C = B^T Q$ and $D=0$ is strictly positive real if $J = -J^T$, $R = R^T >0$ and $Q = Q^T >0$.
\end{corollary}
\begin{Proof}
From Lemma \ref{Lemma:LKY} choose $P = Q$ and {$W_1 = 0$}, then (\ref{Eq:Lemma3}) is trivial, (\ref{Eq:Lemma2}) is $C = B^T Q$ and (\ref{Eq:Lemma1}) becomes
\begin{equation}
\Gamma^T \Gamma = 2QRQ-\varepsilon Q
\end{equation}
then, for $R>0$ there exists a constant $\varepsilon >0$ such that the right hand side is  positive definite, giving a solution for $\Gamma$, using for instance Cholesky factorization. {See Corollary 7.2.9 in \citep{Horn2012Bookmatrix}}. 
\end{Proof}

The next theorem assures the stability of \eqref{Sys:Plant} interconnected in a power preserving way with a SPR controller.
\begin{theo}\label{Theorem:InterconnectionAE}
\cite[Ch.\ 5.1.2]{Villegas2007JournalAport} Consider (\ref{Sys:Plant}) with $u(t)$ defined according to Theorem \ref{Theorem:BCInputs} and $y(t)$ such that
\begin{equation}
\dfrac{1}{2}\dfrac{d}{dt}\Vert z(\ze,t) \Vert_\Lo ^2= u^T(t)y(t),
\end{equation}
{\it i.e.} $W \Sigma W^T=\widetilde{W}\Sigma \widetilde{W}^T=0$ and { $\widetilde{W}\Sigma W^T=I$}.
Consider also a finite-dimensional controller with input $u_c(t)$ and output $y_c(t)$ such that its transfer matrix is SPR. Then, the closed-loop system with the passive interconnection
\begin{subequations}
\begin{align}
u_c (t) &= y(t) \\
u (t) & = -y_c (t)
\end{align}
\end{subequations}
is well-posed and asymptotically stable.
\end{theo}

\subsection{{Equivalent observer based {\it control by interconnection}}.}

In what follows we consider the finite-dimensional approximation \eqref{Sys:DiscretizedPlant} of \eqref{Sys:Plant}.
\begin{definition}{Control by interconnection with a SPR-PH controller.}\label{definition_control_scheme} The considered control scheme is shown in Figure \ref{Fig:PassiveInterconnection}, where the SPR-PH controller is given by
{\begin{equation}\label{Sys:Controller}
\Co\begin{cases}
\dot{\hat{x}}(t) &= (J_c-R_c)Q_c \hat{x}(t) + B_c u_c(t) +B r(t) \\
y_c(t) & = B_c^T Q_c \hat{x} \\
y_r(t) & = B^T Q_c \hat{x},
\end{cases}
\end{equation}
with $J_c=-J_c^T$, $R_c=R_c^T > 0$, $Q_c=Q_c^T > 0 \in \mathbb{R}^{n_c \times n_c}$, $B_c \in \mathbb{R}^{n_c \times n}$ and $B$ defined in (\ref{Sys:DiscretizedPlant}), and (\ref{Sys:Controller}) is interconnected in a power preserving way to the system (\ref{Sys:DiscretizedPlant}), {\it i.e.} such that
\begin{equation} \label{Eq:Interconnection}
u_c (t) = y(t),  \qquad u(t) = r(t)-y_c (t)
\end{equation} 
\begin{figure}[H]
\begin{center}
\includegraphics[height=2.5cm]{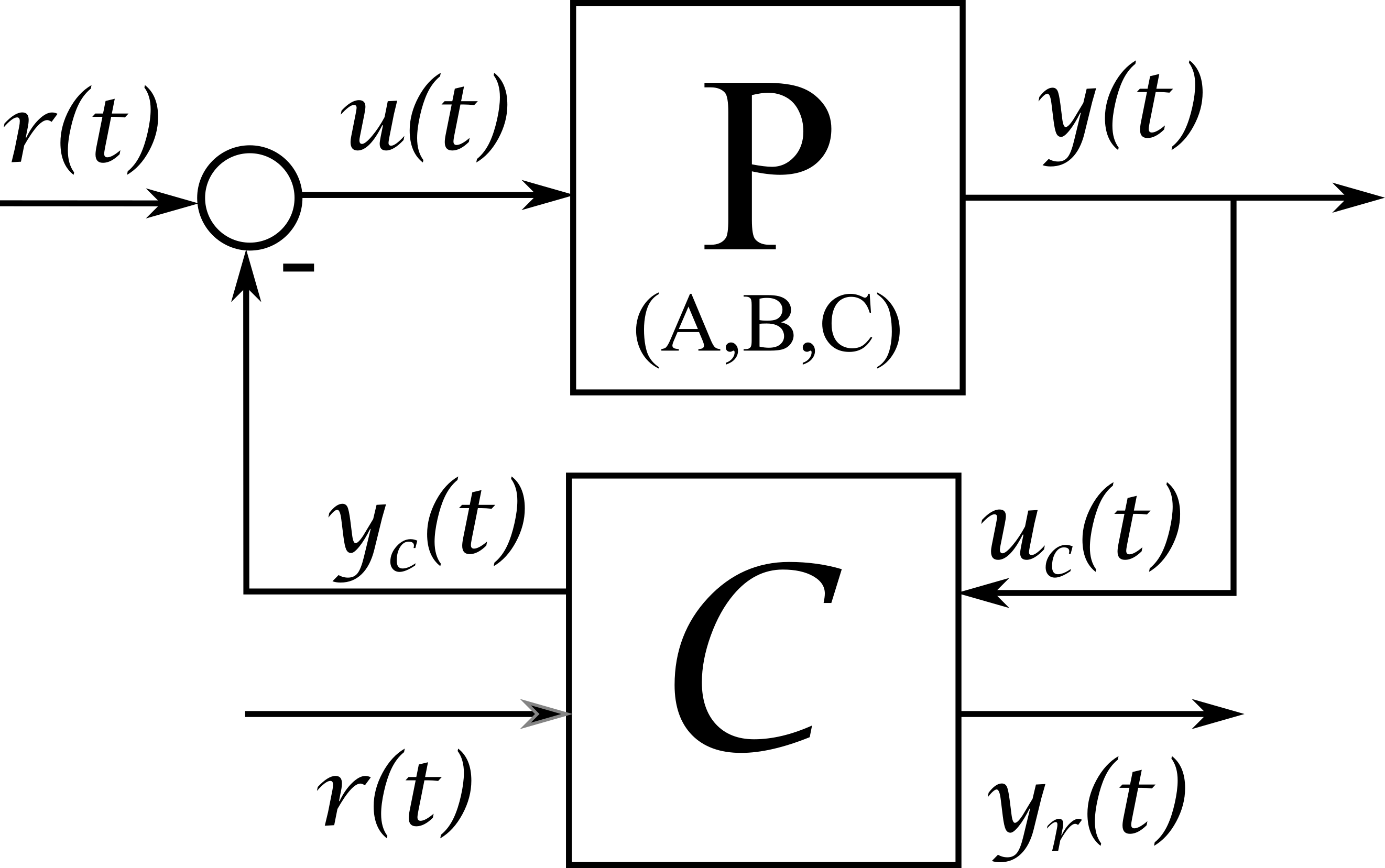}    
\caption{Control scheme}  
\label{Fig:PassiveInterconnection}                                 
\end{center}   
\end{figure}}

Furthermore we assume that $\left( \left(J_c-R_c\right)Q_c,B_c\right)$ is controllable. 
\end{definition}

\begin{theo}\label{Theo:AE_FiniteInterc}
The control scheme of Definition \ref{definition_control_scheme} is asymptotically stable and converges to zero when $r(t)=0$.
\end{theo}
\begin{Proof}
Consider the total energy as Lyapunov function
\[
V(x,\hat{x}) = \dfrac{1}{2} x^T Q x+\dfrac{1}{2} \hat{x}^T Q_c \hat{x}\]
 Then from \eqref{Sys:DiscretizedPlant} and \eqref{Sys:Controller} we have
\[
\dot{V}(x,\hat{x}) \leq - \hat{x}^T Q_cR_cQ_c \hat{x}, \]
where $ R_c>0$ from Definition \ref{definition_control_scheme}. From Lasalle's invariance principle the system converges to the invariant set corresponding to $\dot{V}(x,\hat{x})=0$, {\it i.e.} $\|\hat{x}\|_2=0$. In this case, from \eqref{Sys:Controller}, \eqref{Eq:Interconnection} {and $r=0$ we have} $u_c=y=0$, the controller being controllable. The system \eqref{Sys:DiscretizedPlant} being observable the only equilibrium point is $0$. 
\end{Proof}

In what follows, we propose to start from the achievable closed-loop performances obtained by state feedback to build an observer based controller, {\it i.e.} $J_c$, $R_c$, $Q_c$ and $B_c$ in  \eqref{Sys:Controller} that is strictly positive real, {\it i.e.} satisfies the conditions of Corollary \ref{Corollary:PH_SPR}. 
\begin{proposition} 
\label{Prop:SPR}The interconnection of the system \eqref{Sys:DiscretizedPlant} with the observer based controller \eqref{Eq:ControlLaw}-\eqref{Eq:LuenbergerObserver} is equivalent to {the control by interconnection with the SPR-PH controller of Definition \ref{definition_control_scheme}} if the following matching conditions are satisfied
\begin{equation}\label{Eq:MatchingConditions}
\begin{split}
(J_c-R_c)Q_c &= A-BK-LC \\
B_c^T Q_c & = K \\
B_c & = L.
\end{split} 
\end{equation}
\end{proposition}
\begin{Proof}
The matching equations \eqref{Eq:MatchingConditions} are directly obtained replacing \eqref{Eq:ControlLaw} in \eqref{Eq:LuenbergerObserver} and identifying with \eqref{Sys:Controller} in order to get a passive and  collocated dynamic controller. 
\end{Proof}

\subsection{Main result}

The following proposition is the main contribution of this work. It is based on two main assumptions. 

\begin{assumption}\label{Assump:A_BK}
The matrix $K$ has been designed such that $A-BK$ is Hurwitz by using traditional methods such as LQR design, pole-placement or LMI passivity based control, such as for instance in \citep{Prajna2002JournalAnLMI}.
\end{assumption}

\begin{assumption}\label{Assump:Rc}
The matrix $R_c$ is chosen such that the following matrix
\begin{equation}\label{Eq:HamiltonianMatrix}
H_M = \begin{pmatrix}
A_K & 2R_{c} \\
-C_K & - A_K^T
\end{pmatrix}
\end{equation} with \begin{equation}
A_K = A-BK, \,\,\,\,C_K = -(K^T C + C^T K),
\end{equation}
has no pure imaginary eigenvalues.
\end{assumption}

\begin{remark}
A simple choice for $R_c$ is $R_c = \alpha I$ for some $\alpha>0$ small enough such that the matrix (\ref{Eq:HamiltonianMatrix}) has no pure imaginary eigenvalues.
\end{remark}

\begin{proposition}\label{Prop:Main}
Under Assumptions \ref{Assump:A_BK} and \ref{Assump:Rc}, there exists a matrix $Q_c= Q_c^T >0$, solution of the algebraic Riccati equation (ARE)

\begin{equation}\label{Eq:AREFinal}
A_K^T Q_c + Q_c A_K + 2Q_c R_{c} Q_c + C_K = 0,
\end{equation}

such that the matching equations (\ref{Eq:MatchingConditions}) are satisfied with

\begin{equation}\label{Eq:JcBcL}
\begin{split}
J_c &= \frac{1}{2} \left[ A_K Q_c^{-1} - Q_c^{-1} A_K^T -Q_c^{-1} (K^T C-C^T K)  Q_c^{-1} \right] \\
B_c &= Q_c^{-1}K^T \\
L &= B_c.
\end{split}
\end{equation}

Furthermore, the matrix $A-LC$ is Hurwitz.
\end{proposition}

\begin{Proof}
From \citep{Kosmidou2007JournalGeneralized} it is known that if the Hamiltonian matrix (\ref{Eq:HamiltonianMatrix}) has no pure imaginary eigenvalues then there exists a solution $Q_c = Q_c^T >0$ for (\ref{Eq:AREFinal}). Hence we only need to prove that (\ref{Eq:AREFinal}) is compatible with the matching equation (\ref{Eq:MatchingConditions}) for $J_c$ and $L$ as in (\ref{Eq:JcBcL}). Since $Q_c$ is invertible and solution of (\ref{Eq:AREFinal}) we have

\begin{equation}\label{Eq:Rc}
\begin{split}
R_c &= -\frac{1}{2}\left[Q_c^{-1} A_K^T + A_KQ_c^{-1} + Q_c^{-1}C_K  Q_c^{-1}\right]  \\
	&= -\frac{1}{2}\left[Q_c^{-1} A_K^T + A_KQ_c^{-1} - Q_c^{-1}(K^T C + C^T K)  Q_c^{-1}\right]  \\
\end{split}
\end{equation} 

Then using (\ref{Eq:JcBcL}) and (\ref{Eq:Rc}) we have
\begin{equation}
\begin{split}
(J_c-R_c)Q_c &= \frac{1}{2}   (2 A_K Q_c^{-1} -2 Q_c^{-1} K^T C Q_c^{-1}) Q_c \\
			&=  A_K  - Q_c^{-1} K^T C  \\
			&=  A_K  - L C  \\
			&=  A-BK  - L C  \\
\end{split}
\end{equation}
which correspond to (\ref{Eq:MatchingConditions}). From Theorem \ref{Theo:AE_FiniteInterc} the closed-loop system
\begin{equation}\label{Eq:AugmentedSys}
\dfrac{d}{dt}\begin{pmatrix}
x \\
\hat{x}
\end{pmatrix} = \begin{pmatrix}
A & -BK \\
B_c C & (J_c-R_c)Q_c
\end{pmatrix}\begin{pmatrix}
x \\
\hat{x}
\end{pmatrix}+ \begin{pmatrix}
B \\ 
B
\end{pmatrix}r
\end{equation}
is asymptotically stable. Applying the following transformation
\begin{equation*}
\begin{pmatrix}
x\\
\tilde{x}
\end{pmatrix}= \begin{pmatrix}
I & 0 \\
I & -I
\end{pmatrix}\begin{pmatrix}
x\\
\hat{x}
\end{pmatrix}
\end{equation*}
the closed-loop system (\ref{Eq:AugmentedSys}) can be written
\begin{equation}\label{Eq:AugmentedSys2}
\begin{split}
\dfrac{d}{dt}\begin{pmatrix}
x \\
\tilde{x}
\end{pmatrix} = &\begin{pmatrix}
A_K & BK \\
A_K-B_cC-A_c & A_c+BK
\end{pmatrix}\begin{pmatrix}
x \\
\tilde{x}
\end{pmatrix}\\  & \,\,+ \begin{pmatrix}
B \\ 
B
\end{pmatrix}r \\
\end{split}
\end{equation}
with $A_K= A-BK$, $B_c = L$ and $A_c=(J_c-R_c)Q_c = A-BK-LC$ or equivalently
\begin{equation}\label{Eq:AugmentedSys3}
\dfrac{d}{dt}\begin{pmatrix}
x \\
\tilde{x}
\end{pmatrix} = \begin{pmatrix}
A-BK & BK \\
0 & A-LC
\end{pmatrix}\begin{pmatrix}
x \\
\tilde{x}
\end{pmatrix}+ \begin{pmatrix}
B \\ 
0
\end{pmatrix}r
\end{equation}
Since $A_K$ is Hurwitz, and the closed-loop system asymptotically stable,  $A-LC$ is also Hurwitz.
\end{Proof}

\begin{theo} Let's consider the infinite-dimensional system \eqref{Sys:Plant}
with $u=-K \hat{x}$ and $\hat{x}$ solution of the dynamic equation \eqref{Sys:Controller} in Proposition \ref{Prop:Main}. The closed-loop system is asymptotically stable. \end{theo}
{\begin{Proof} 
The proof is a direct application of Theorem \ref{Theorem:InterconnectionAE}, Proposition \ref{Prop:SPR} and Proposition \ref{Prop:Main}.  Choosing $R_c>0$ satisfying Assumption \ref{Assump:Rc} from Proposition \ref{Prop:Main} there exists a matrix $Q_c=Q_c^T >0$ solution of \eqref{Eq:AREFinal} such that the finite-dimensional observer defined by \eqref{Eq:LuenbergerObserver} is stable and the matching equations \eqref{Eq:MatchingConditions} are satisfied. {Then,} from Proposition \ref{Prop:SPR} the control is equivalent to the control by interconnection with a SPR-PH controller. From Theorem \ref{Theorem:InterconnectionAE} the closed-loop system is well posed and asymptotically stable when the finite-dimensional controller is applied to the infinite-dimensional system.
\end{Proof}}

\begin{remark}
One special case of Proposition \ref{Prop:Main} is proposed in \citep{Wu2018JournalReduced} in the context of reduced order control of finite-dimensional PHS. There the matrix $K$ obtained by a LQR method and the matrix $Q_c=Q$.
\end{remark}
\section{Example: the Timoshenko beam} \label{Section:Example}

We consider the boundary control of a Timoshenko beam clamped at the left side and controlled through force and torque at the right side. Both longitudinal  and angular velocities at the right side are used for control purposes. {The port Hamilonian formulation of the Timoshenko beam can be found in \citep{Macchelli2004JournalModeling}}. It can be written in the form (\ref{Sys:Plant}) with $$P_1=\left[ \begin{array}{cccc}0 & 1 & 0& 0 \\ 1 & 0 & 0& 0 \\ 0 & 0 & 0& 1 \\ 0 & 0 & 1& 0  \end{array}\right], \; P_0=\left[ \begin{array}{cccc}0 & 0 & 0& -1 \\ 0 & 0 & 0& 0 \\ 0 & 0 & 0& 0 \\ 1 & 0 & 0& 0  \end{array}\right],$$
$$\Lo (\ze)=\left[ \begin{array}{cccc}
T(\ze) & 0 & 0& 0 \\
0 & \frac{1}{\rho(\ze)} & 0& 0 \\
0 & 0 & EI(\ze)& 0 \\
0 & 0 & 0& \frac{1}{I_\rho (\ze)}  \end{array}\right],G_0=0$$
where $T(\ze)$ is the shear modulus, $\rho(\ze)$ the mass per length unit, $EI(\ze)$ the product of the Young's modulus of elasticity $E$ and the moment of inertia of a cross section $I$, and $I_\rho(\ze)$  the moment of inertia of a cross section.

The state variables are: the shear displacement, the transverse momentum distribution, the angular displacement and the angular momentum distribution defined respectively by $z_1(\ze,t)= \tfrac{\partial w}{\partial \ze}(\ze,t) - \phi(\ze,t)$, $z_2(\ze,t)= \rho(\ze)\tfrac{\partial w}{\partial t}(\ze,t)$, $z_3(\ze,t) = \tfrac{\partial \phi}{\partial \ze}(\ze,t)$ and $z_4(\ze,t)= I_\rho (\ze)\tfrac{\partial \phi}{\partial t}(\ze,t)$, where $w(\ze,t)$ and $\phi(\ze,t)$ are respectively the transverse displacement of the beam and the rotation angle of a neutral fiber of the beam. Note that, $T(\ze)z_1(\ze,t)$ is the shear force,$\frac{1}{\rho(\ze)} z_2(\ze,t)$ the longitudinal velocity, $EI(\ze) z_3(\ze,t)$  the torque and $ \frac{1}{I_\rho (\ze)} z_4(\ze,t)$ the angular velocity.

We choose as inputs and outputs
\begin{equation*}\label{Eq:InputsOutputs}
u(t)=\begin{pmatrix}
\frac{1}{\rho(a)} z_2(a,t) \\
\frac{1}{I_\rho (a)} z_4(a,t) \\
T(b) z_1(b,t) \\
EI(b) z_3(b,t) \\
\end{pmatrix}, \,\, y(t)=\begin{pmatrix}
-T(a) z_1(a,t) \\
-EI(a) z_3(a,t) \\
\frac{1}{\rho (b)} z_2(b,t) \\
\frac{1}{I_\rho(b)} z_4(b,t) \\
\end{pmatrix}
\end{equation*}
 
The total energy of the beam is defined as
\begin{equation*}
H(t) = \frac{1}{2} \Vert z(\ze,t) \Vert_\Lo ^2 = \frac{1}{2} \int_a^b z^T(\ze,t) \Lo (\ze) z(\ze,t) d\ze.
\end{equation*}
and satisfies
\begin{equation*}
\frac{d}{dt}H(t) = y^T(t)u(t).
\end{equation*}
\subsection{Discretization}
The infinite-dimensional system is discretized using finite differences on staggered grids \citep{Trenchant2018JournalFinite} {considering 20 elements per state variable, which leads to $n_c=80$.} The finite-dimensional model \eqref{Sys:DiscretizedPlant} is then given by
\begin{equation*} J = \left[ \begin{array}{cccc}
0 & D & 0 & -F \\
-D^T & 0 & 0 & 0 \\
0 & 0 & 0 & D \\
F^T & 0 & -D^T & 0 \\
\end{array}\right]
\end{equation*}
\begin{equation*}
 R = 0, \,\,\,\, Q = h\left[ \begin{array}{cccc}
Q_1 & 0 & 0 & 0 \\
0 & Q_2 & 0 & 0 \\
0 & 0 & Q_3 & 0 \\
0 & 0 & 0 & Q_4 \\
\end{array}\right]
\end{equation*}
where $Q_i, \; i \in \left\{1,\cdots,4\right\}$ are diagonal matrices containing the evaluation of  $T(\ze)$, $\frac{1}{\rho(\ze)}$, $EI(\ze)$ and $\frac{1}{I_\rho(\ze)}$ respectively, at the specific points chosen for the discretization.
\begin{equation*}
D=\dfrac{1}{h^2}\left[ 
\arraycolsep=3pt\def\arraystretch{0.9}
\begin{array}{cccc}
1 & 0 & \cdots & 0 \\
-1 & 1 & \ddots & 0  \\
\vdots & \ddots & \ddots & \ddots \\
0 & 0 & \cdots & 1 \\
\end{array}\right], \;
F=\dfrac{1}{2h}\left[ 
\arraycolsep=3pt\def\arraystretch{0.9}
\begin{array}{cccc}
1 & 0 & \cdots & 0 \\
1 & 1 & \ddots & 0  \\
\vdots & \ddots & \ddots & \ddots \\
0 & 0 & \cdots & 1 \\
\end{array}\right],
\end{equation*}

and
\begin{equation*}
B = \left[ \begin{array}{cccc}
b_{11} & b_{12} & 0 & 0 \\
0 & 0 & b_{23} & 0 \\
0 & b_{32} & 0 & 0 \\
0 & 0 & b_{43} & b_{44} \\
\end{array}\right]
\end{equation*}

with
\begin{equation*}
b_{11}=\dfrac{1}{h}\left[ \begin{array}{c}
-1 \\
0 \\
\vdots  \\
0 \\
\end{array}\right], \,\,\,\,
b_{12}=\dfrac{1}{2}\left[ \begin{array}{c}
-1 \\
0 \\
\vdots  \\
0 \\
\end{array}\right],
\end{equation*}
\begin{equation*}
b_{23}=\dfrac{1}{h}\left[ \begin{array}{c}
0 \\
0 \\
\vdots  \\
1 \\
\end{array}\right], \,\,\,\, b_{43}=\dfrac{1}{2}\left[ \begin{array}{c}
0 \\
0 \\
\vdots  \\
1 \\
\end{array}\right]
\end{equation*}
\begin{equation*}
b_{32} = b_{11}, \,\,\,\, b_{44}=b_{23}.
\end{equation*}
The state variables are
$$x(t)=\left[ \begin{array}{cccc}
x_1^d(t)^T & x_2^d(t)^T & x_3^d(t)^T & x_4^d(t)^T \end{array}\right]^T$$
where $x_i^d(t)\in \mathbb{R}^{20}$, $i \in \left\{ 1,\cdots,4\right\}$ {and the $j-th$ component of $x_1^d$, $x_2^d$, $x_3^d$ and $x_4^d$ correspond respectively to the approximation of $z_1((j-0.5)h,t)$, $z_2(jh,t)$, $z_3((j-0.5)h,t)$ and $z_4(jh,t)$, with $h= 2\tfrac{b-a}{2*20+1} = 0.0146$, and $b-a = 0.3 \; m$ with $a = 0$.} The beam is clamped at the left side and force and torque actuators at the right side are considered, $T z_1(b,t)$ and $EI z_3(b,t)$ respectively. Hence $b_{11} = b_{12} = b_{32} = 0$, which give pairs $(A,B)$ controllable and $(A,C)$ observable. In this case $C = B^T Q$.

\subsection{The {observer-based} state feedback design}\label{Subsection:Design}
{We use the parameters of Table \ref{Table:param}.
\begin{table}[H]
\center
\begin{tabular}{|l|l|l|}
\hline
Parameters & Values & Unit \\\hline
$T $& $3.4531\times 10^5 $&$ Pa$ \\
$\rho$ & $0.0643 $&$ kg.m^{-1}$ \\
EI & $37.0116 $&$ Pa.m^4$ \\
$I_\rho$ & $2.1485\times 10^{-6} $&$ Kg.m^2$\\
 $ [a,b]$ & $[0,0.3] $&$m$\\ \hline
\end{tabular}
\caption{Plant parameters.\label{Table:param}}
\end{table}}

\begin{figure}
\begin{center}
\includegraphics[width=0.48\textwidth ]{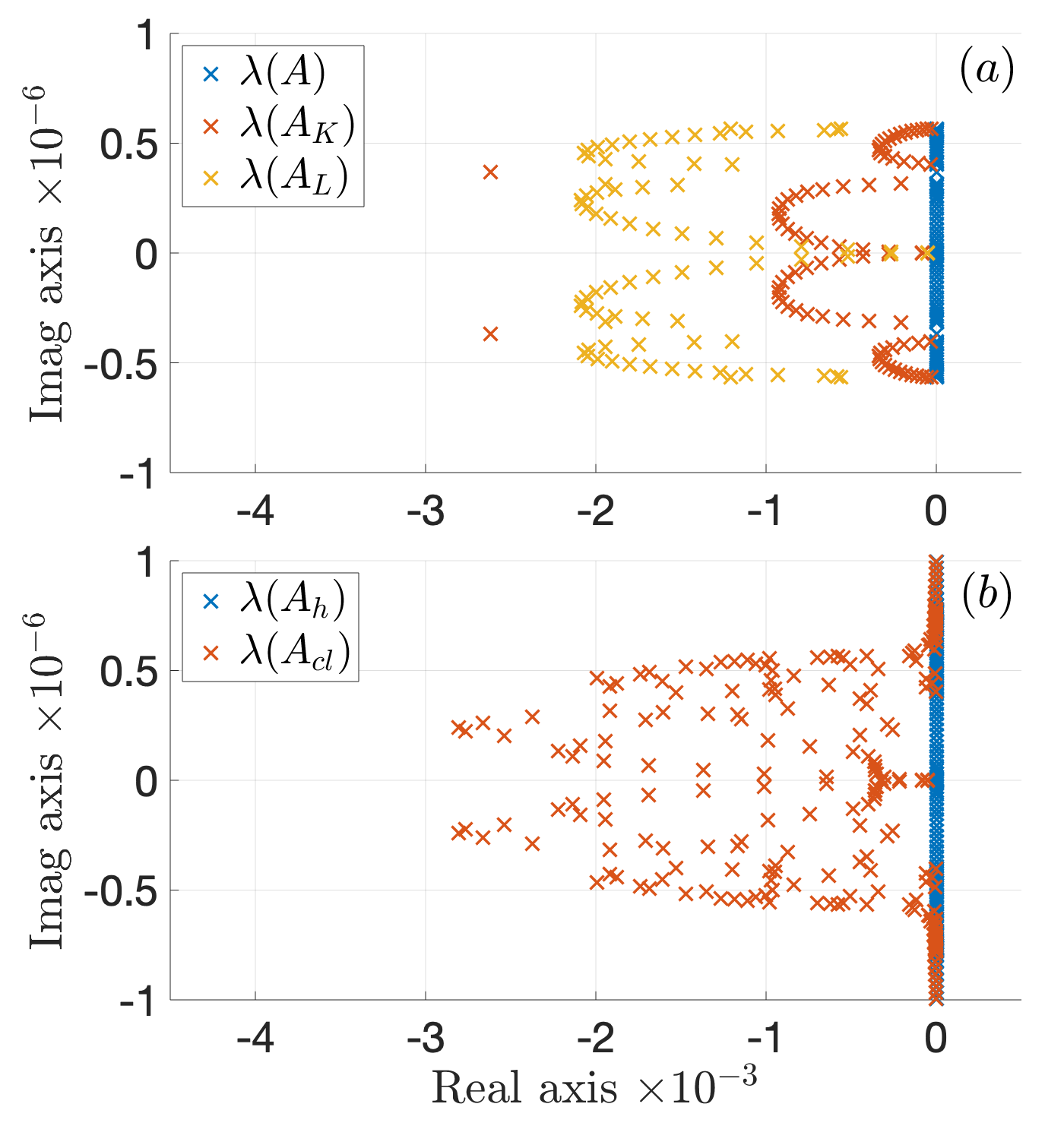}    
\caption{(a): $\lambda (A)$: eigenvalues of the discretized model with {$n_c=80$}, $\lambda (A_K)$: eigenvalues of $A-BK$  and $\lambda (A_L)$: eigenvalues of  $A-LC$. {(b): $\lambda (A_h)$: eigenvalues of the discretized model with $n_c=200$}, $\lambda (A_{cl})$: closed-loop eigenvalues.}  
\label{Fig:EigenvaluesAllAREDesign2}                                 
\end{center}                                 
\end{figure}

{Two different state feedbacks $K$ minimizing the cost function 
 \[ J_{LQR} = \int_0^\infty {\lbrace x^T Q_{LQR}x + u^T R_{LQR} u + 2x^T N_{LQR} u \rbrace }dt \] 
 are designed using the Matlab@ Control System Toolbox {\it lqr.m}. In both designs the ARE algorithm proposed in \citep{Lanzon2008JournalComputing} has been used to solve \eqref{Eq:AREFinal}.
 
For the first design, the matrix $K$ is performed choosing $Q_{LQR}=0.8I_{n_c}$, $R_{LQR}=10I_{4}$ and $N_{LQR}=0$, while the matrix $L$ is designed following Proposition \ref{Prop:Main} with $R_c=10I_{n_c}$. The eigenvalues of the matrices $A$, $A-BK$ and $A-LC$ are shown in Figure \ref{Fig:EigenvaluesAllAREDesign1} (a). The eigenvalues of the closed-loop system using the same controller on a higher order discretization of the beam, choosing  {$n_c = 200$}, are given in  Figure \ref{Fig:EigenvaluesAllAREDesign1} (b).

\begin{figure}
\begin{center}
\includegraphics[width=0.48\textwidth]{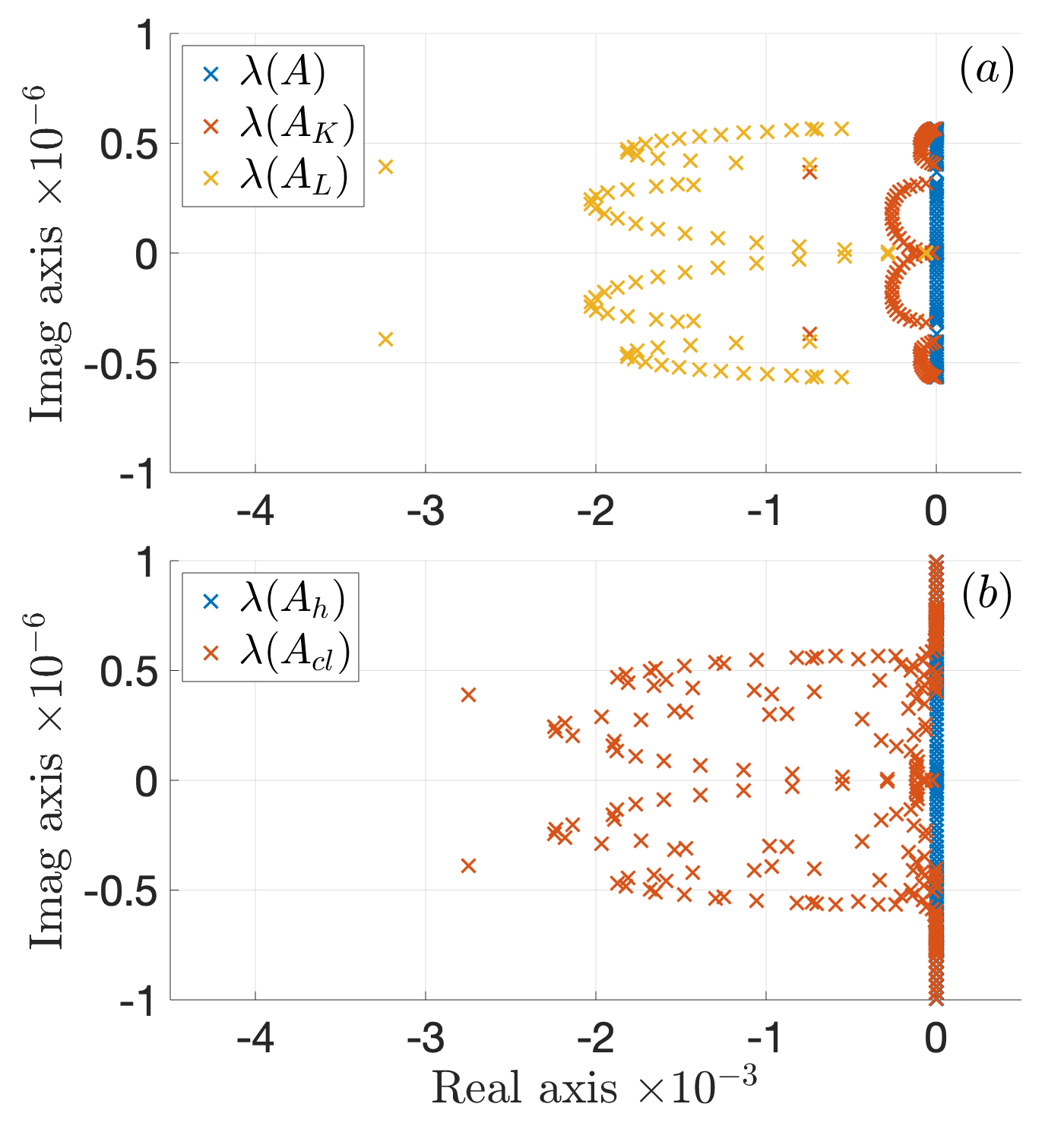}    
\caption{(a): $\lambda (A)$: eigenvalues of the discretized model with {$n_c=80$}, $\lambda (A_K)$: eigenvalues of  $A-BK$ and $\lambda (A_L)$: eigenvalues of $A-LC$. {(b): $\lambda (A_h)$: discretized model eigenvalues with $n_c=200$}, $\lambda (A_{cl})$: closed-loop eigenvalues.}  
\label{Fig:EigenvaluesAllAREDesign1}                                 
\end{center}                                 
\end{figure}

For the second design, the matrix $K$ is performed choosing $Q_{LQR}=0.8I_{n_c}$, $R_{LQR}=1.33I_{4}$ and $N_{LQR}=0$, while the matrix $L$ is designed following Proposition \ref{Prop:Main} with $R_c=4I_{n_c}$. The eigenvalues of the matrices $A$, $A-BK$ and $A-LC$ are shown in Figure \ref{Fig:EigenvaluesAllAREDesign2} (a). The eigenvalues of the closed-loop system using the same controller on a higher order discretization of the beam, choosing  {$n_c = 200$}, are given in  Figure \ref{Fig:EigenvaluesAllAREDesign2} (b). In both cases, the closed-loop system remains stable and the high frequency modes are not destabilized. Even if $n_c \rightarrow \infty$, the closed-loop eigenvalues do not cross the imaginary axis.}

\subsection{Simulations}
\begin{figure}[H]
\begin{center}
\includegraphics[width=0.48\textwidth]{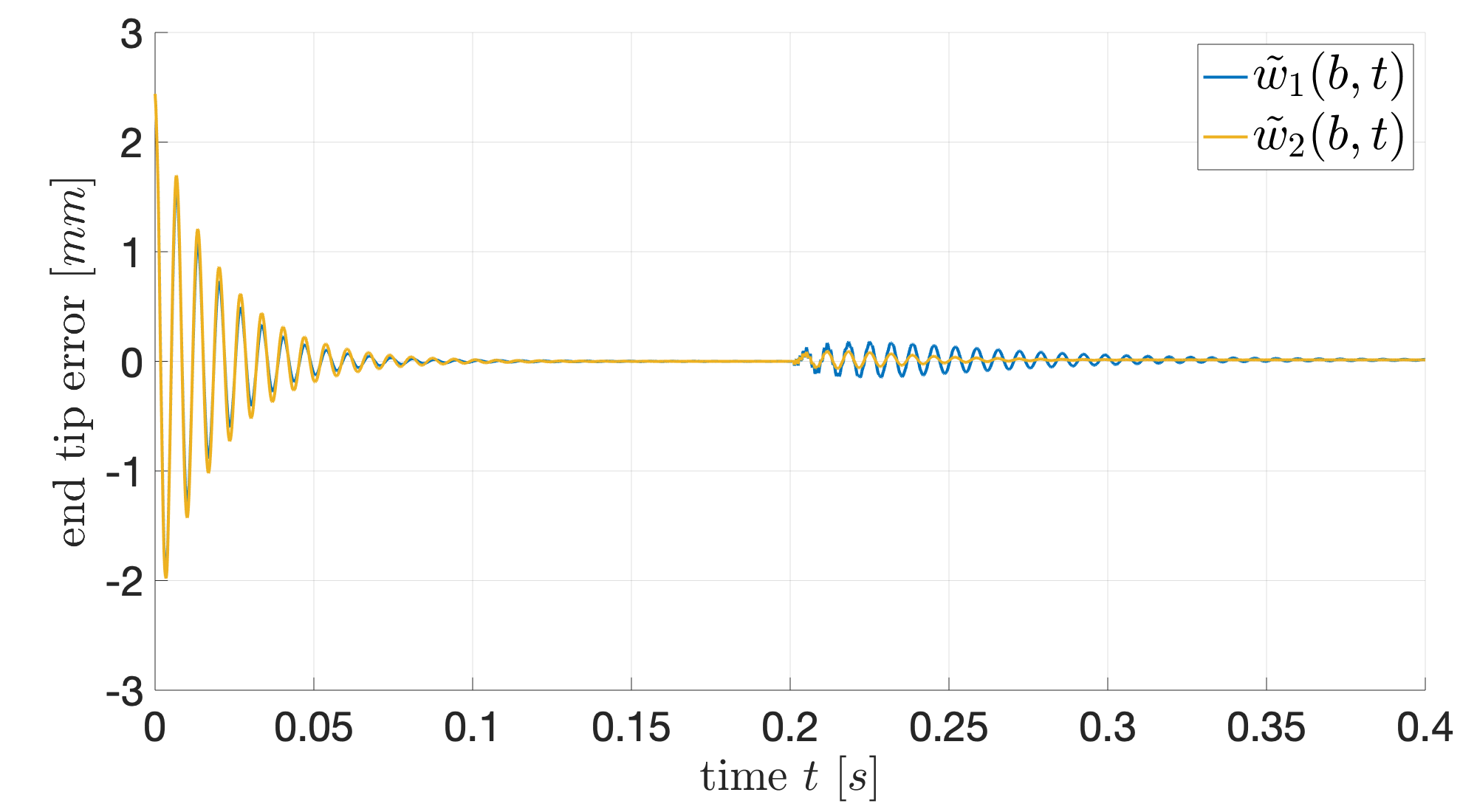}    
\caption{{$\tilde{w}_i(b,t)$: observer error of the end tip displacement for the design $i = \lbrace 1,2 \rbrace$.}}  
\label{Fig:ObserverDesign}                                 
\end{center}
\end{figure}

\begin{figure}[H]
\begin{center}
\includegraphics[width=0.48\textwidth]{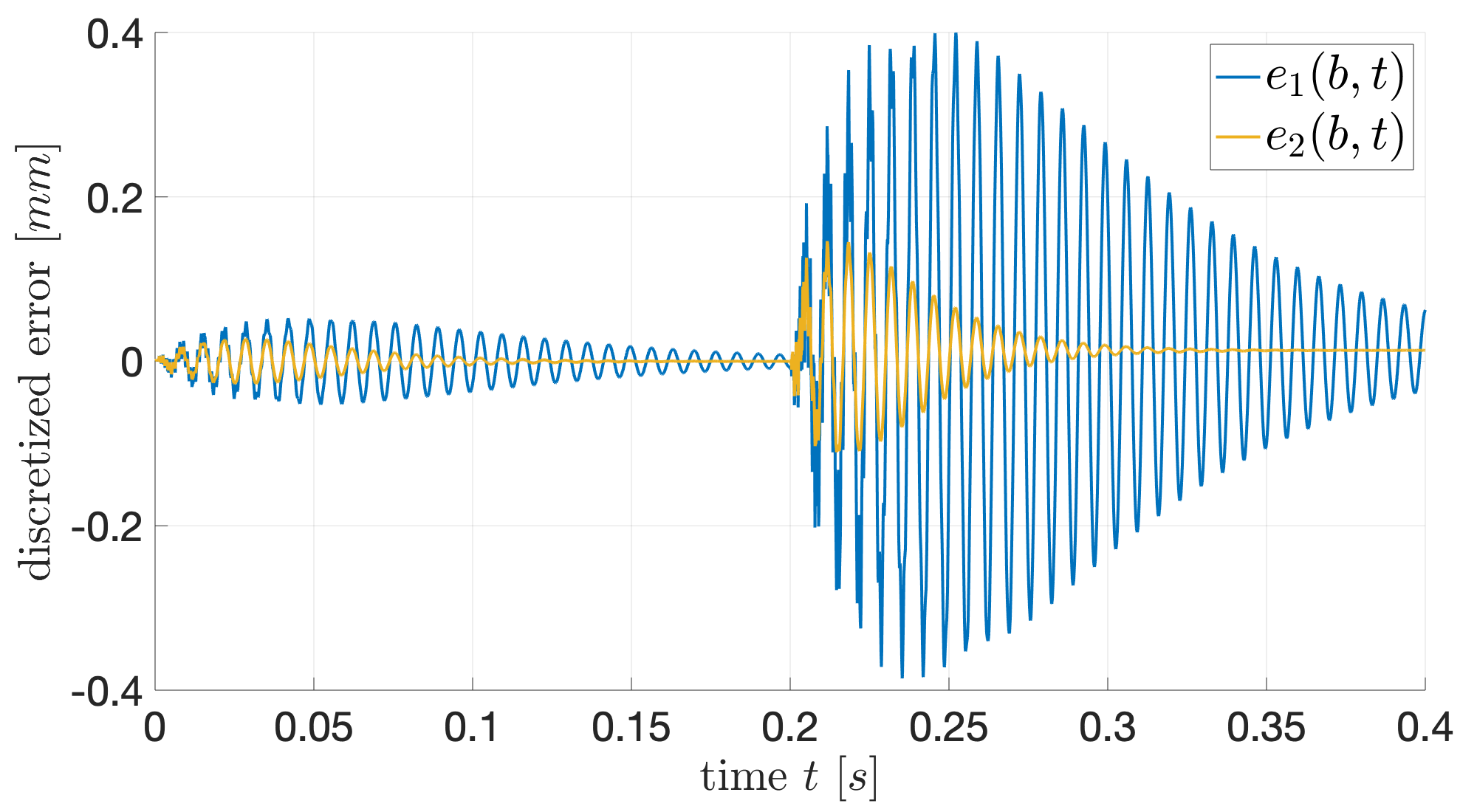}    
\caption{{Discretization error $e_i(b,t) = w_i^\Po(b,t)-w_i^P(b,t)$ for the design $i= \lbrace 1,2 \rbrace$. }}  
\label{Fig:RealVsDiscretized}                                 
\end{center}
\end{figure}

{Simulations are performed using Matlab over the time interval $t\in[0,0.4] \; s$. The simulation starts from the initial condition $z_1(\ze,0) = 0.2896 \times 10^{-4}$, $z_2(\ze,0) = 0$, $z_3(\ze,0) = -0.2702 \ze +0.0811$ and $z_4(\ze,0) = 0$ corresponding to the equilibrium position associated to a force of $10N$ applied at the end tip of the beam. The initial condition for the observer is set to zero, i.e. $\hat{x}(0) = 0$. An external force $r(t)=100 N$ is applied at $t=0.2s$ at the end of the tip to modify the equilibrium position. In the first simulation the {observer-based} controller of size 80 and designed on the discretized model for 20 elements is applied to the large scale system obtained considering 50 discretization elements, {\it i.e.} 200 state varibles. Figure \ref{Fig:DifferentDesign} shows the time responses for the two controllers proposed in subsection \ref{Subsection:Design}, where $w_1(b,t)$ and $w_2(b,t)$ are the end tip displacements of the beam for the first and second design respectively. Note that, before $t = 0.2s$ the convergence to the null equilibrium is due to the observer and state feedback dynamics. After the step at $t = 0.2s$, the convergence is mostly due to the state feedback as the observer already converged to the system state over the considered range of frequencies.

\begin{figure}[H]
\begin{center}
\includegraphics[width=0.48\textwidth]{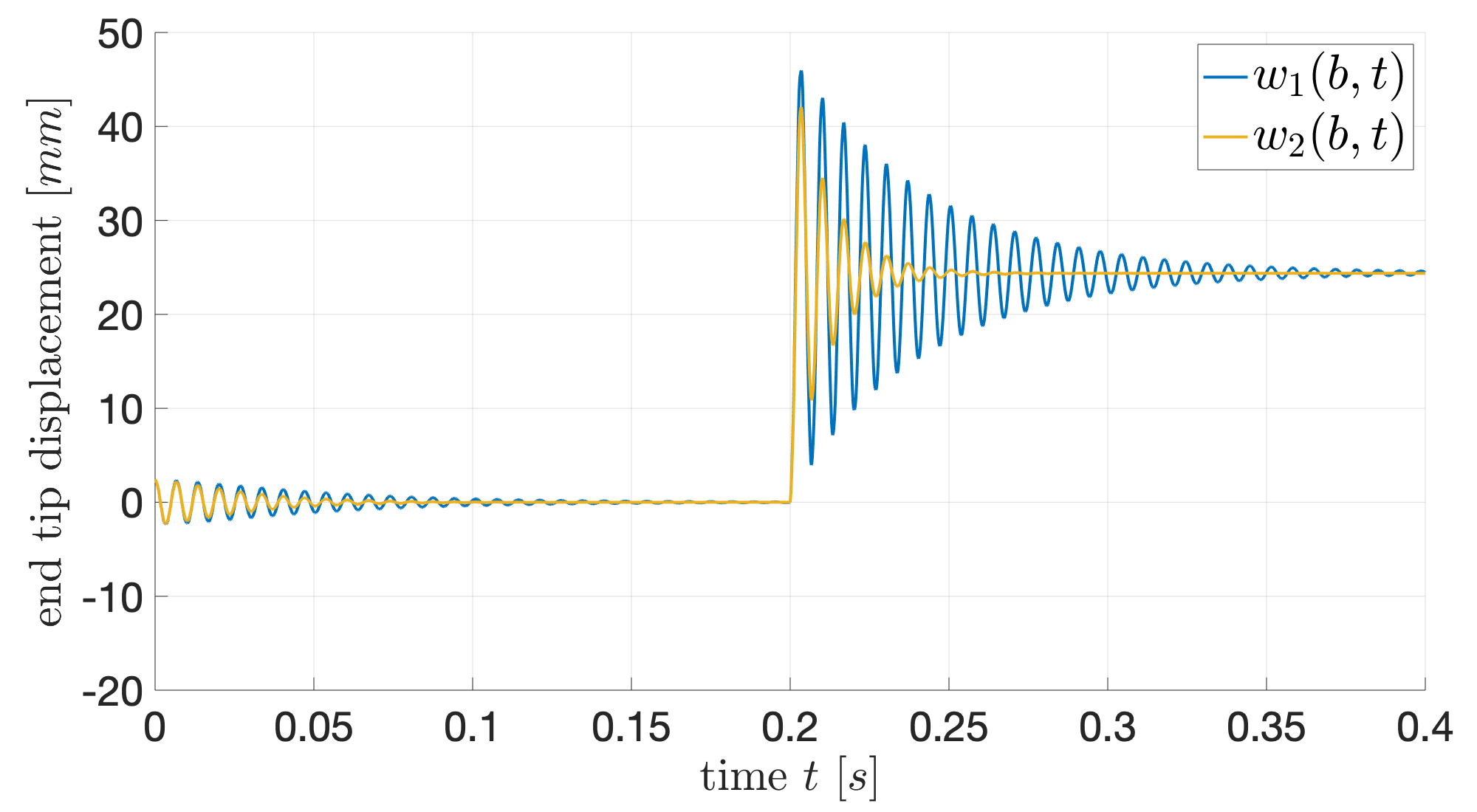}    
\caption{{$w_i(b,t)$: end tip displacement for the design $i = \lbrace 1,2 \rbrace$.}}  
\label{Fig:DifferentDesign}                                 
\end{center}
\end{figure}

The estimated values are given by $\hat{w}_1(b,t)$ and $\hat{w}_2(b,t)$, the error is obtained by $\tilde{w}_i(b,t) = w_i(b,t)-\hat{w}_i(b,t)$ and it is shown in Figure \ref{Fig:ObserverDesign} for both designs.}
{In Figure \ref{Fig:ErrorNorm} is given the evolution of the norm of the error between the system state, that has been used for control design, and the observer state, with respect to time when considering different initial conditions. Figure \ref{Fig:ErrorNorm} illustrates the convergence rate of the observer on the reduced order system. 
\begin{figure}[H]
\begin{center}
\includegraphics[width=0.48\textwidth]{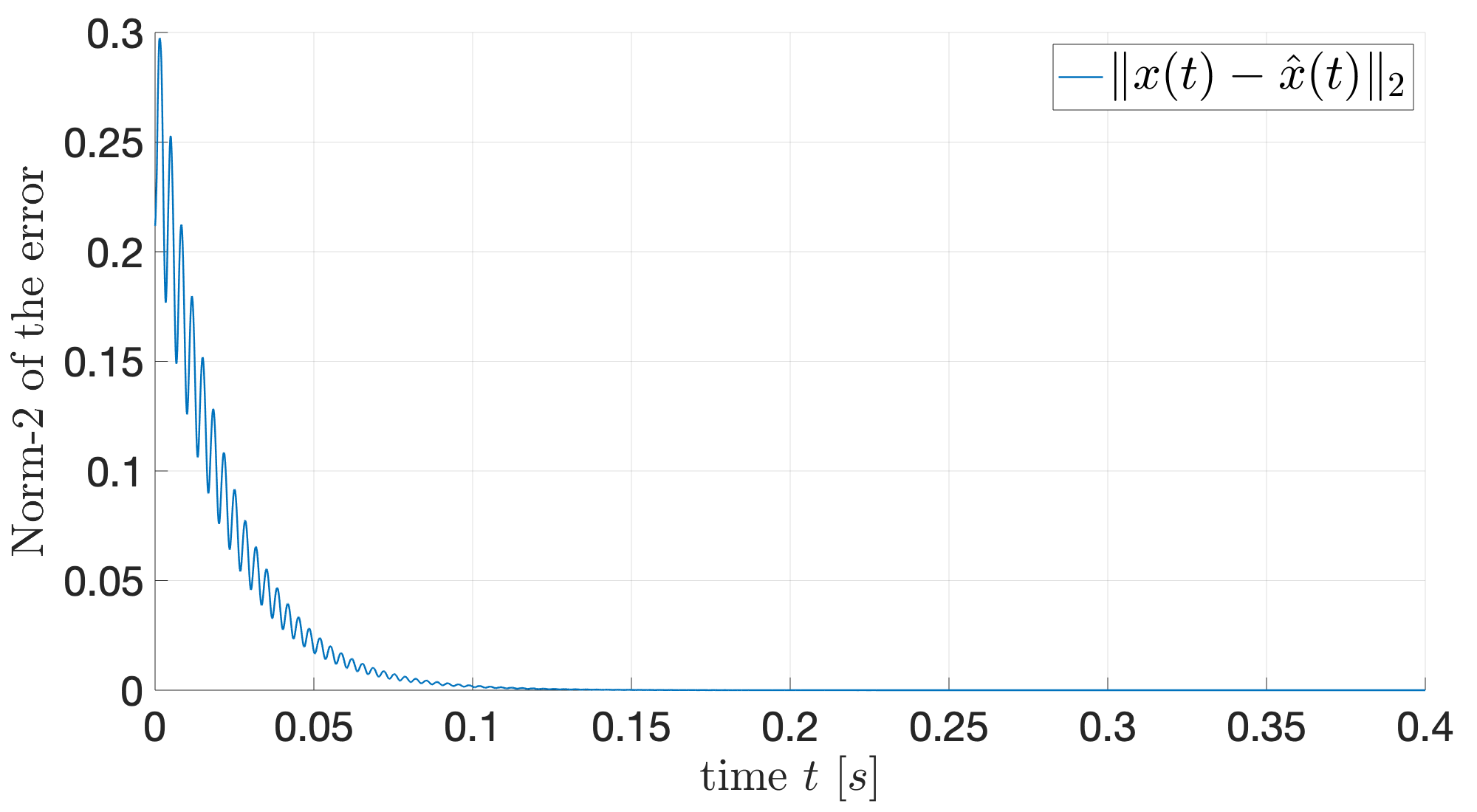}    
\caption{Evolution of the norm of the error between the observer and the discretized model to initial conditions.}  
\label{Fig:ErrorNorm}                                 
\end{center}
\end{figure}}
{Since the controller is designed based on a finite-dimensional approximation $P$ of the system, but at the end, has to be implemented on the infinite-dimensional system $\Po$, it is interesting to compare the behavior of both closed-loop systems. For that purpose, we denote by $w^P(b,t)$ the end tip displacement when applying the controller to the finite-dimensional model $P$ used for the design (20 elements) and by $w^\Po(b,t)$ the end tip displacement when applying the controller to a higher dimensional model stemming from a fine approximation of the infinite-dimensional model $\Po$ (obtained for 50 elements). The {\it approximation error} $e_i(b,t) = w_i^\Po(b,t)-w_i^P(b,t)$  is shown in Figure \ref{Fig:RealVsDiscretized} for both designs. We can notice that the {\it approximation error} increases for high frequencies signals, meaning performances cannot be guaranteed over all frequencies. Yet the error remains small with respect to the higher order approximation. }

{Finally, a new simulation is done using the second design for $t \in [0.0.2]$ with the same initial conditions than before and an external force step applied at $t = 0.1s.$ The deformation of the beam along the space and over time is shown in Figure \ref{Fig:Deformation}. 
The oscillations ocurring during the first $0.1 s$ are due to the observer convergence since the system and the observer do not have the same initial conditions.   
\begin{figure}[H]
\begin{center}
\includegraphics[width=0.48\textwidth]{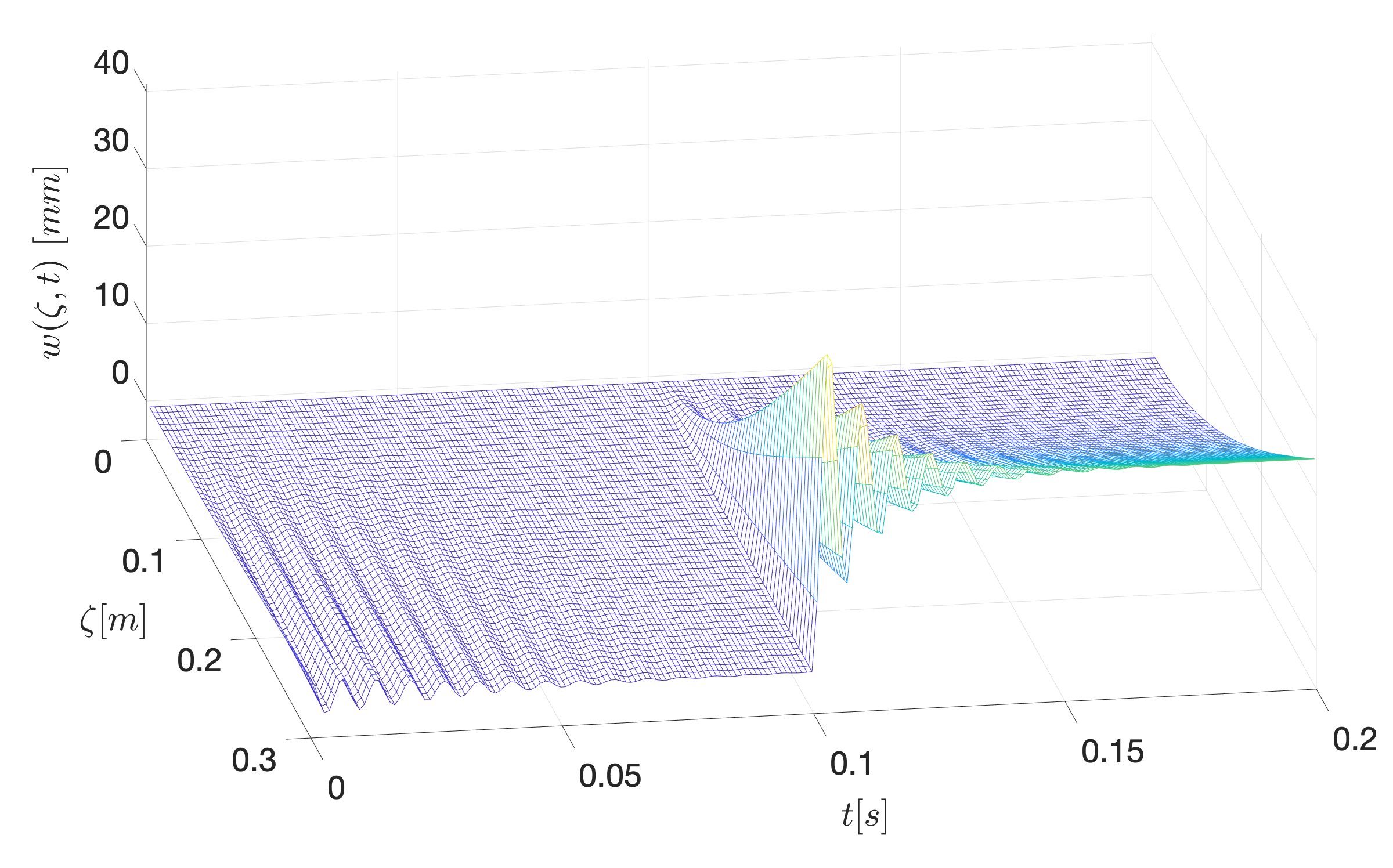}    
\caption{Displacement of the beam for time $t \in [0,0.2]$ and space $\ze \in [0,0.3]$.}  
\label{Fig:Deformation}                                 
\end{center}
\end{figure}

Figure \ref{Fig:EstimatedEnergies} shows the evolution of the system and observer energies with respect to time when considering different initial conditions. We can see that the observer energy function converges to the plant energy function, and at the same time the control brings the closed-loop energy function to zero. 
\begin{figure}[H]
\begin{center}
\includegraphics[width=0.48\textwidth]{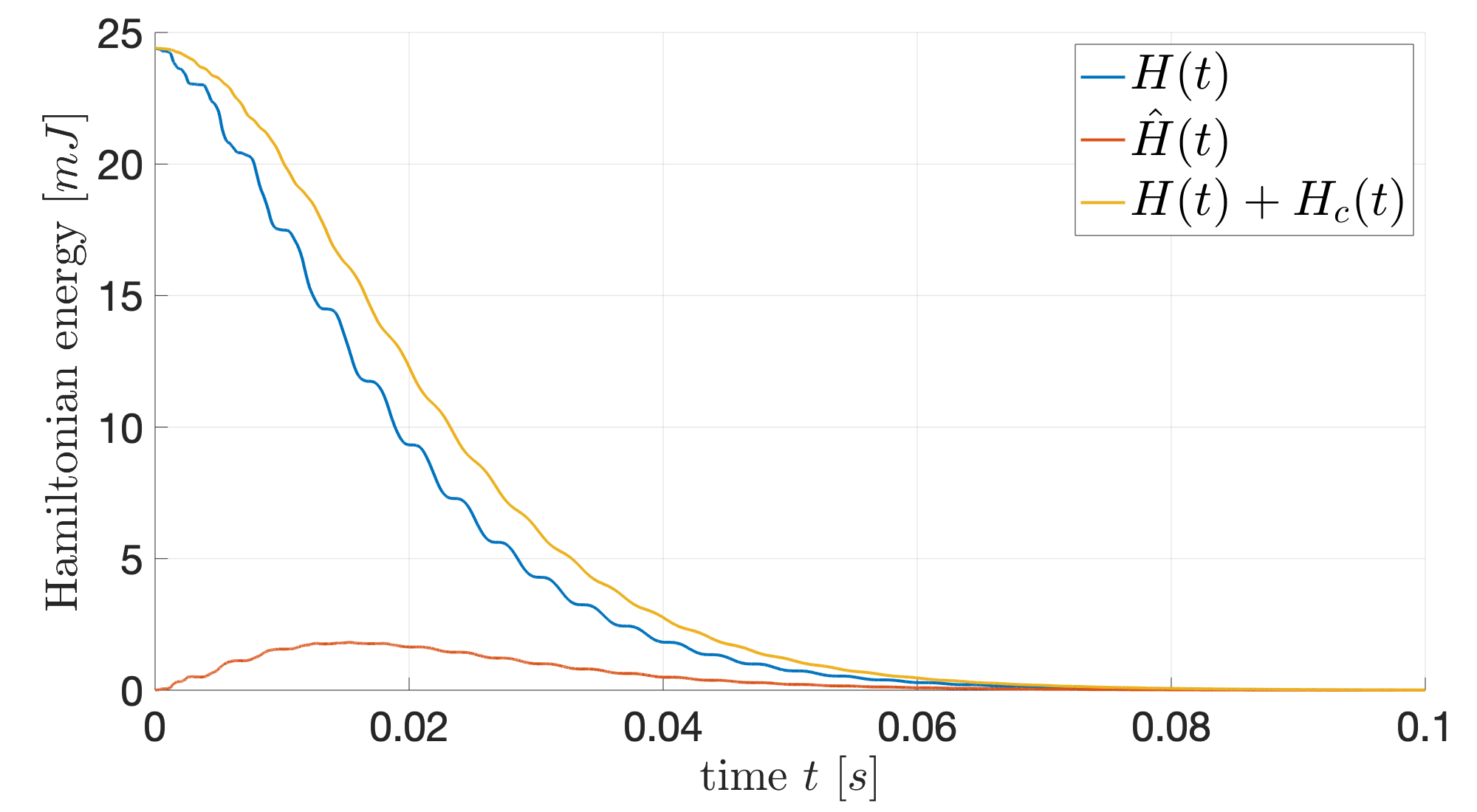}    
\caption{Evolution of the system and observer energy functions with respect to initial conditions.}  
\label{Fig:EstimatedEnergies}                                 
\end{center}
\end{figure}
}

\section{Conclusion} \label{Section:Conclusions}
An observer based boundary controller has been proposed for a class of boundary controlled PHS defined on a 1D spatial domain. The design is based on an early-lumping approach in which a finite-dimensional PHS approximation of the infinite-dimensional system is used to design the observer and the controller. The main contribution is a constructive method that guarantees that the finite-dimensional dynamic boundary controller is a strictly positive real PHS. This guarantees that the interconnection between the controller and the infinite-dimensional system is asymptotically stable. {As soon as the finite-dimensional approximation of the system that is used for the observer design is close to the infinite-dimensional system over the considered range of frequencies, the closed-loop performances on the infinite-dimensional system are close to the ones obtained on the finite-dimensional approximation.} The stabilization of a Timoshenko beam with force and torque actuators and collocated measurements (velocities) has been used to illustrate the approach.

\section*{Acknowledgements}
This work has been supported by the French-German ANR-DFG INFIDHEM project ANR-16-CE92-0028 and the the EIPHI Graduate School (contract ANR-17-EURE-0002).  The second author has received founding from Bourgogne-Franche-comt\'e Region ANER 2018Y-06145. The third author acknowledges Chilean FONDECYT 1191544 and CONICYT BASAL FB0008 projects. The fourth author has received funding from the European Unions Horizon 2020 research and innovation programme under the Marie Sklodowska-Curie grant agreement No 765579.

\bibliographystyle{elsarticle-harv}
\biboptions{authoryear}
\bibliography{bibliographie}

\end{document}